\documentclass[twocolumn,letterpaper]{IEEEAerospaceCLS}

\usepackage{amsmath}
\usepackage{amssymb}
\usepackage{graphicx}
\usepackage{color}
\usepackage{commath}
\usepackage{gensymb}
\usepackage{algorithm}
\usepackage{algpseudocode}
\usepackage{booktabs} 
\usepackage[colorlinks=true, urlcolor=blue, linkcolor=black, citecolor=black]{hyperref}

\DeclareMathOperator*{\minimize}{minimize}
\DeclareMathOperator*{\subjectto}{subject\ to}

\begin{document}

\title{Inter-Satellite Link Configuration for Fast Delivery in Low-Earth-Orbit Constellations}

\author{
Arman Mollakhani\\
Northwestern University\\
Dept. of Electrical and Computer Engineering\\
2145 Sheridan Road, Evanston, IL 60208\\
arman.mollakhani@northwestern.edu
\and
Jerayu Tiamraj\\
Northwestern University\\
Dept. of Electrical and Computer Engineering\\
2145 Sheridan Road, Evanston, IL 60208\\
tomtiamraj2025@u.northwestern.edu
\and
Shu-Jie Cao\\
Northwestern University\\
Dept. of Electrical and Computer Engineering\\
2145 Sheridan Road, Evanston, IL 60208\\
shujie.cao@northwestern.edu
\and
Dongning Guo\\
Northwestern University\\
Dept. of Electrical and Computer Engineering\\
2145 Sheridan Road, Evanston, IL 60208\\
dguo@northwestern.edu
}

\maketitle

\thispagestyle{plain}
\pagestyle{plain}

\begin{abstract}
End-to-end latency in large low-Earth-orbit (LEO) constellations is dominated by propagation delay, making total delay roughly proportional to the network diameter---the longest shortest-path in hops.  
Current inter-satellite link (ISL) layouts have rarely been optimized to minimize network diameter while 
simultaneously satisfying physical and operational constraints, including maximum link distance, line-of-sight, per-satellite hardware limits, and long-term link viability over orbital periods. 
In this study, the selection and assignment of inter-plane ISLs is formulated as a diameter-minimization problem on a Starlink-inspired Walker–Delta constellation in which each satellite is equipped with two fixed intra-plane links and may activate up to two inter-plane links. 
Beginning with a feasible baseline, the topology is iteratively refined by a local-search procedure that replaces or reinforces links to shrink the diameter.
The resulting ISL configuration meets all geometric and hardware limits, preserves link stability across multiple orbital periods, and yields a sparse, diameter-aware graph with potential for centralized routing capabilities. Simulations demonstrate
the proposed algorithm achieves low worst-case latency
without compromising ISL stability, and the trade-off between hop count and long-term link stability is empirically measured for guidance of future LEO network deployments.

\end{abstract}

\tableofcontents

\section{Introduction}\label{Introduction}

Recent low-Earth-orbit (LEO) constellations such as Starlink~\cite{fcc21}, OneWeb~\cite{fcc20oneweb}, Amazon's Kuiper~\cite{fcc20kuiper}, and AST SpaceMobile~\cite{fcc25ast_mod} aim to provide global connectivity by providing worldwide coverage. 
Well-configured optical inter-satellite links (ISLs) can enhance the efficiency of these constellations, making them appealing
for latency-critical services like real-time remote control, remote surgery and telecare, high-frequency trading, blockchain synchronization, and emergency alerting~\cite{bozkurt2017why,mollakhani2025fault}. In free space, light travels approximately 300 km per millisecond
(covering the geodesic distance between the North and South Poles in about 66.7 ms), roughly
47\% faster 
than in fiber~\cite{handley2018delay}.
Hence, a well-designed
LEO ISL network
can potentially surpass the fastest terrestrial cables on intercontinental routes. Achieving such capabilities requires that the worst-case end-to-end latency be tightly bounded.

Meeting these latency requirements relies
on 
the unique geometry of LEO constellations.
Geostationary Earth Orbit (GEO) satellites, positioned at approximately \(35{,}786\;\text{km}\), incur round-trip propagation delays of
hundreds of milliseconds, far too high for interactive or
low-latency applications. 
In contrast, LEO satellites,
orbiting below \(2{,}000\;\text{km}\), enable  
tightly interconnected optical meshes
with per-hop propagation delays
an order of magnitude smaller than those in
GEO constellations. At the same time, LEO networks' launch cost and transmit power requirements are more favorable than those of medium-Earth-orbit (MEO) systems.
This
positions LEO constellations as the premier platform for
low-latency global networking in the foreseeable future~\cite{ntrs_2010_mooring_nasa,berry2024spectrum}.

Unlike terrestrial networks that use distributed protocols such as Internet Protocol (IP) and Border Gateway Protocol (BGP), LEO constellations operated by a single service provider, such as Starlink, can employ
centralized routing strategies. Routes and forwarding tables can be pre-computed for short \emph{snapshots} during which satellite positions and candidate links are treated as static, then pushed to all nodes before the next snapshot begins~\cite{handley2018delay}. This determinism eliminates the need for in-orbit route discovery and enables cut-through switching~\cite{kermani1979virtual}, or multiprotocol label switching (MPLS)-style label forwarding~\cite{rfc3031}, allowing each packet to be steered with microsecond-scale processing overhead per hop~\cite{heine2023end}.
Moreover, if 
ISL capacity
is
provisioned to handle the offered traffic demands, queuing delays at each node can also be constrained to the microsecond scale with high probability. Under these conditions, where per-hop processing delay is negligible compared to propagation delay,
end-to-end latency is dominated by the cumulative propagation delay across all traversed links. 
Since each hop introduces an additional propagation segment, the total delay grows approximately in proportion to the number of hops taken along the path. While minimizing the absolute path distance would be ideal, we instead use the network diameter---the largest shortest-path cost between any two satellites---as a structural proxy for latency. While a general problem is formulated using arbitrary (positive) link cost, we use hop count as the specific measure for path cost to solve the optimization problem in simulations. This choice is justified because LEO constellations form sparse geographical graphs (each satellite has only a few inter-satellite links), and link lengths are upper-bounded by orbital geometry. Consequently, a lower hop count typically implies a straighter, more direct route, reducing cumulative propagation distance and avoiding inefficient zig-zagging across the network. 

Reducing the diameter is straightforward in principle: add more inter-plane ISLs and choose them so that distant orbits are bridged early in a path.  In practice, however, each satellite is limited to a small number of steerable laser terminals, and any candidate link must satisfy distance and line-of-sight constraints not just at one instant but ideally throughout an orbital period.  The design problem is therefore to select a sparse, physically feasible set of inter-plane ISLs that keeps the diameter small while remaining stable over time.
This paper
addresses this challenge for a Walker–Delta constellation
where each satellite maintains two fixed intra-plane links and may activate up to two inter-plane links (as in Starlink)~\cite{fcc21,wang1993structural}. The ISL-selection task is cast as a constrained diameter-minimization problem, and an iterative local-search algorithm is 
proposed to refine an initial feasible topology.  Extensive simulations show that the resulting configurations (i) respect all geometric and hardware limits, (ii) remain viable across multiple orbital periods, and (iii) reduce worst-case hop count and thus latency.  By quantifying the trade-off between hop reduction and long-term link stability, the study offers concrete guidance for next-generation LEO network deployments.

The main contributions of this paper are:
\begin{enumerate}
    \item Formulating 
    the inter-plane ISL selection
    as a constrained network diameter minimization problem for a realistic Walker-Delta constellation, considering hardware, geometric, and link stability constraints.
    
    \item Proposing
    an iterative local-search algorithm that begins with a feasible baseline topology and systematically refines it by replacing and reinforcing links to reduce network diameter while ensuring
    long-term viability.

    \item Demonstrating, through extensive simulation,
    that the optimized topologies yield substantial reductions in worst-case latency, along with
    the first empirical analysis of the trade-off between minimum hop count and long-term link stability, offering practical guidance for future LEO network designs.
\end{enumerate}

The rest of this paper is organized as follows. Section~\ref{s:related} surveys prior work on ISL topology design. Section~\ref{s:model} describes the constellation model and physical constraints. Section~\ref{Problem Formulation} formalizes the diameter-minimization problem under two different operational models. The proposed iterative optimization algorithm is detailed in Section~\ref{Inter-Satellite Link Topology Design}. Simulation results are presented in Section~\ref{Simulation & Results}, and Section~\ref{Conclusion} concludes the paper with directions for future research.

\section{Related Work}
\label{s:related}
The use of large-scale LEO satellite constellations for global broadband coverage is proposed as a promising direction for enabling low-latency, scalable communication. In such systems, satellites are deployed in dense orbital configurations and interconnected through ISLs to form a mesh-like network~\cite{handley2018delay}. It is argued that the effectiveness of these constellations depends heavily on how the underlying inter-satellite topology is designed, with efficient link assignment and routing considered essential for supporting time-sensitive applications.

Several strategies for ISL topology design have been explored in prior work. A time-division-based method for ISL topology generation has been proposed, in which different sets of feasible links are periodically activated to maintain long-term connectivity~\cite{chu2018time}. Focusing on the ISL assignment problem for navigation constellations, link scheduling has been optimized for time synchronization and autonomous orbit determination. To solve this multi-objective problem under payload and visibility constraints, assignment algorithms based on satellite layering and weighting within a discrete-time system of superframes have been developed~\cite{sun2022inter}. To tackle the computational complexity of the large-scale ISL scheduling problem, it has been modeled as a time-discrete network multi-commodity flow problem, and a data-driven parallel adaptive large neighborhood search (DP--ALNS) algorithm has been proposed to find solutions that minimize total data transmission delay~\cite{liu2022data}. These approaches emphasize temporal reachability but do not explicitly minimize worst-case broadcast latency or network diameter. 

Multicast and broadcast strategies in LEO constellations have also been explored, including core-based multicast trees and disruption-tolerant mechanisms for multilayered satellite networks~\cite{cheng2007core,zheng2010routing}. The broader challenge of routing in LEO networks has been extensively surveyed, with trade-offs identified between centralized and distributed protocols and an emphasis placed on the need for algorithms that can adapt to predictable topological changes~\cite{xiaogang2016survey}. Building on the predictable topology of LEO constellations, a software-defined networking (SDN) based approach has been introduced in which a centralized terrestrial controller pre-computes routing tables for all anticipated topologies. These routing tables are then distributed to space nodes in advance, enabling them to switch based on time or network context, thereby eliminating in-network routing convergence time and reducing the computational load on satellites~\cite{corici2024sdn}. While these methods improve resilience and group communication, they often rely on idealized routing assumptions or neglect geometric constraints inherent to LEO dynamics.

Other efforts have focused on low-latency path design under stochastic satellite distributions. Tools from stochastic geometry have been used to model spatial connectivity and evaluate average-case latency~\cite{wang2022stochastic}, and congestion-aware routing portocols have been developed to improve throughput under dynamic traffic conditions~\cite{dai2021distributed}. However, these approaches generally target statistical performance rather than deterministic guarantees or worst-case latency bounds.

Further analysis of path selection has introduced a theoretical model for explicitly estimating the ISL hop-count between ground users, showing how the hop-count is determined by user latitudes and longitude differences~\cite{chen2021analysis}. An explicit analytic algorithm was later proposed to solve the shortest distance path (SDP) problem by modeling it with satellite phase and converting it into a total phase offset problem that can be solved directly without graph-based iterations. It has also been proven that almost all SDPs lie within the minimum hop path (MHP) set, thereby simplifying the search space for optimal routing~\cite{chen2024shortest}.

While these studies cover a broad range of routing algorithms and dynamic link scheduling strategies, the fundamental problem of designing a sparse and stable ISL topology to explicitly minimize the worst-case latency remains an open challenge. Our work addresses this gap by focusing on diameter minimization under realistic physical constraints.

\section{System Model}
\label{s:model}

In this work, we adopt a snapshot-based view of the LEO satellite network, where the topology is assumed to remain static within each snapshot and only updated at discrete time intervals. Each snapshot represents a fixed configuration of satellite positions and feasible ISLs, during which routing and link assignment decisions can be made deterministically. 
This formulation allows for tractable analysis and static optimization of network properties within each snapshot, motivating our snapshot-level topology optimization framework that seeks to minimize worst-case latency under physical and structural constraints.
A key requirement of this formulation is that each selected ISL remains viable throughout the snapshot duration, meaning it must be geometrically feasible, free of Earth obstruction, and within allowable distance bounds for the entire snapshot duration. By designing sparse yet viable ISL configurations that ensure bounded worst-case hop 
counts across the network, the method provides a practical framework for enabling fast global message dissemination in real-world LEO satellite deployments.
 
We consider an LEO satellite constellation arranged in a Walker-Delta configuration with $N_p$ orbital planes and $N_s$ satellites per plane, for a total of $N = N_p \times N_s$ satellites.
Each satellite orbits at a circular altitude of $h$ km above Earth's surface with an inclination of $\theta\degree$. The orbital radius is $r=R_E + h$, where $R_E = 6371$ km is the mean radius of Earth. Furthermore, a satellite is associated with a unique position in three-dimensional Cartesian space denoted by $x_u \in \mathbb{R}^3$, where $u \in \{1,\dots,N\}$ is the satellite index. Satellites are grouped into orbital planes $P_1,P_2,\dots,P_{N_p}$, with each plane $P_i$ containing $N_s$ satellites.
The maximum feasible distance for an ISL is denoted by $d_{max}$, determined by
physical-layer constraints.

Orbital planes are evenly spaced in longitude, and the satellites within each plane are uniformly distributed. To introduce spatial asymmetry, a random angular phase offset $\phi_i \in [0,\phi_{\max}]$ is applied to each plane $i$. This preserves the regular constellation structure while capturing practical position variations.

We represent the satellite constellation as an undirected graph $G=(V,E)$, where the vertex set $V$ contains all $N$ satellites, and the edge set $E$ represents the collection of all active ISLs.
Each satellite is assumed to be equipped with high-speed, bidirectional optical terminals for inter-satellite communication. The network includes two types of links:
\begin{itemize}
    \item \textbf{Intra-plane Links {($E_{\text{intra}}$)}:} Each satellite maintains bidirectional connections with its nearest neighbors in the same orbital plane, forming a fixed ring topology.
    \item \textbf{Inter-plane Links {($E_{\text{inter}}$)}:} Each satellite {$v \in V$} can establish up to $L_{\text{inter}}$ {bi}directional links with satellites in different orbital planes, subject to physical feasibility. Let $\text{deg}_{\text{inter}}(v)$ be the degree of vertex $v$ considering only edges in $E_{\text{inter}}$; this is constrained such that:
    \begin{align}
    \deg_{\text{inter}}(v) &\leq L_{\text{inter}} \quad \forall v \in V. \label{eq:inter_degree_bound}
    \end{align}
\end{itemize}

The edge set is a union of fixed intra-plane links, $E_{\text{intra}}$, and selectable inter-plane links, $E_{\text{inter}}$, such that $E = E_{\text{intra}} \cup E_{\text{inter}}$.
A candidate inter-plane link between two satellites $u \in P_i$ and $v \in P_j$
is considered feasible if the following conditions are met:

\begin{enumerate}
    \item Their Euclidean distance {must not exceed a maximum link distance $d_{\max}$}:
    \begin{align}
        \|x_u - x_v\| \leq d_{\max}. \label{eq:distance_constraint}
    \end{align}
    \item The line-of-sight between satellites must not obstructed by Earth, that is,   
    \begin{align}
        \frac{\| x_u \times x_v \|}{\| x_u - x_v \|} > R_E \label{eq:los_constraint}
    \end{align}
    where ``$\times$'' denotes the cross product of two position vectors.
\end{enumerate}

We model the network using a static snapshot of satellite positions. This approach allows routes to be precomputed for each epoch, enabling deterministic path planning and reducing in-network operational overhead. Such a pre-planned environment is ideally suited for low-latency forwarding techniques like cut-through switching and MPLS.
In cut-through switching, a packet is forwarded to the next satellite as soon as the destination address is read, without the full packet being received~\cite{kermani1979virtual}.
This minimization of buffering delay is particularly effective in high-speed optical networks where propagation delay dominates. Similarly, MPLS enables efficient packet forwarding by attaching short labels to packets, allowing each satellite to determine the next hop based on precomputed paths rather than full network lookups~\cite{rfc3031}. These features align naturally with the centralized control and predictable topology of LEO satellite networks.

Building on this snapshot-based framework, we consider two models for inter-plane link feasibility that reflect different operational assumptions. In the first, a link is considered feasible if it satisfies geometric constraints, namely distance and line-of-sight, at a single snapshot in time. This snapshot-only model offers greater flexibility in link selection but does not guarantee long-term physical validity as satellites move along their orbits. To address, this we consider a viability-constrained model that requires a link to remain feasible throughout multiple orbital cycles. This stricter condition yields topologies that are more robust and operationally deployable, as they reduce the need for frequent link reconfiguration.
This distinction is particularly important in practice, as reconfiguring optical ISLs is non-trivial. Switching from one target to another requires physically steering laser terminals, re-acquiring beam alignment, and establishing a stable optical connection. Recent studies have shown that the setup time for such links can be on the order of several seconds, due to the delays associated with beam acquisition and pointing adjustments~\cite{bhattacharjee2023laser}. 
These delays introduce non-negligible communication gaps and limit the feasibility of frequent link switching. As a result, topologies that remain valid over extended time periods are preferred for maintaining stable and low-latency performance in LEO satellite networks.

\section{Problem Formulation}
\label{Problem Formulation}

Our objective is to select the set of inter-plane links $E_{\text{inter}}$ that minimizes the worst-case communication cost across the constellation. We define the 
shortest-path cost between two satellites $u$ and $v$ as $d_G(u,v)$. This cost metric can be modeled in several ways. In a \textbf{weighted graph model}, each link can be assigned a weight representing physical latency components, such as propagation delay (proportional to distance), transmission delay (based on link capacity), or even dynamic factors like queuing delay. In this case, $d_G(u,v)$ would be the minimum sum of weights along paths from $u$ to $v$.

Alternatively, in a simpler \textbf{unweighted model}, each link is assigned a uniform cost of one, making $d_G(u,v)$ the shortest-path hop distance. Given that LEO constellations are sparse graphs with bounded link lengths, hop count serves as a practical and effective proxy for total latency.

The objective is to minimize the network diameter, defined as:
\begin{align}
    \operatorname{diam}(G) = \max_{u,v \in V} d_G(u,v). \label{eq:diameter}
\end{align}

We formulate the optimization problem under two different operational assumptions. In the flexible \textbf{snapshot-only} model, link feasibility is assessed at a single instant in time. The set of all possible inter-plane links that satisfy the geometric constraints \eqref{eq:distance_constraint} and \eqref{eq:los_constraint} at that snapshot is denoted by $\mathcal{E}_{\text{potential}}$. 
The optimization problem is formally stated as:
\begin{subequations}\label{eq:optimization_problem}
\begin{align}
    \minimize_{E_{\text{inter}}} \quad
    & \max_{u, v \in V}  d_G(u,v) \label{eq:objective}\\
    \subjectto \quad 
    & \text{deg}_{\text{inter}}(v) \leq L_{\text{inter}}, && \forall v \in V \label{eq:degree_constraint} \\
    & \|x_u - x_v\| \leq d_{\max}, && \forall (u,v) \in E_{\text{inter}} \label{eq:dist_feasibility} \\
    & \frac{\| x_u \times x_v \|}{\| x_u - x_v \|} > R_E, && \forall (u,v) \in E_{\text{inter}} \label{eq:los_feasibility}
\end{align}
\end{subequations}

In contrast, the more robust \textbf{viability-constrained} model imposes a stricter requirement that links must remain feasible over an extended duration, such as a full orbital period $T$, ensuring the topology is stable. Let $x_u(t)$ be the position of satellite $u$ at time $t$. An edge $(u,v)$ is included in the smaller potential set
only if it satisfies the geometric constraints for the entire duration.
This yields a smaller feasible set than $
\mathcal{E}_{\text{potential}}$, trading off some potential optimality for improved robustness and long-term deployability. The optimization problem for the viability-constrained model is formally stated as:
\begin{subequations}\label{eq:optimization_problem_viable}
\begin{align}
    \minimize_{E_{\text{inter}}} & \quad \max_{u, v \in V} \quad d_G(u,v) \label{eq:objective_viable}\\
    \subjectto & \quad \text{deg}_{\text{inter}}(v) \leq L_{\text{inter}}, \quad \forall v \in V \label{eq:degree_constraint_viable} \\
    & \quad \|x_u(t) - x_v(t)\| \leq d_{\max}, \label{eq:dist_feasibility_viable} \\
    & \qquad \forall (u,v) \in E_{\text{inter}}, \; \forall t \in [t_0, t_0+T] \nonumber \\
    & \quad \frac{\| x_u(t) \times x_v(t) \|}{\| x_u(t) - x_v(t) \|} > R_E, \label{eq:los_feasibility_viable}\\
    & \qquad \forall (u,v) \in E_{\text{inter}}, \; \forall t \in [t_0, t_0+T] \nonumber
\end{align}
\end{subequations}

Finding a globally optimal solution to the diameter minimization problem formulated above is computationally prohibitive for a large-scale constellation. Consequently, this work proposes an effective heuristic algorithm based on iterative local search to construct a low-diameter, physically feasible ISL topology. The proposed methodology is detailed in the following section.

\section{ISL Topology Design}
\label{Inter-Satellite Link Topology Design}

The construction of the ISL topology proceeds in two phases.
First, an initial feasible network graph is generated based on the constellation's geometry and hardware constraints. Second, this baseline topology is iteratively optimized using a local search algorithm to reduce the network diameter.

\subsection{Initial Topology Construction}\label{Initial Topology Construction}
We construct a baseline network graph using a static snapshot of satellite positions, as described in Section~\ref{s:model}. Each satellite $u \in P_i$ is connected to its two intra-plane neighbors in the same orbital plane $P_i$, forming a fixed ring topology.
For inter-plane links, feasible candidate connections are determined based on the geometric constraints introduced earlier: (i) a maximum allowable link distance $d_{\max}$, and (ii) unobstructed line-of-sight. 

To construct a sparse but viable inter-plane topology, the algorithm proceeds as follows:
\begin{itemize}
    \item For each satellite, all feasible inter-plane link candidates are identified based on their position and visibility.
    \item Satellites with fewer inter-plane links are prioritized for new connections, promoting a more balanced degree distribution across the constellation.
    \item For a given satellite, its candidate links are strategically ordered: after sorting in ascending order, candidates are split into the farthest and nearest halves, each half is shuffled, and the shuffled farthest-half candidates are considered first.
    \item  A link is established with the first valid candidate in this sequence, provided the candidate satellite has not yet reached its own inter-plane link limit, $L_{\text{inter}}$, and the link does not already exist. This process is repeated multiple times to ensure that as many satellites as possible are assigned their full quota of inter-plane links.
\end{itemize}

As described in Section~\ref{s:model}, this initial topology does not enforce long-term link viability. However, once constructed, the graph is later evaluated for orbital stability over time, allowing us to assess the trade-off between link longevity and structural performance and providing us with the option to enforce this in the next step.

The resulting sparse configuration, composed of static intra-plane links and selected inter-plane connections, serves as the initial state for the iterative optimization process described in the next subsection.

\subsection{Iterative Topology Optimization}\label{Iterative Topology Optimization}
 After the initial graph is constructed, 
an iterative local search algorithm refines the inter-plane connections to reduce the network's diameter, defined in~\eqref{eq:diameter}. The optimization alternates between two phases: periodic reinforcement to correct under-linked satellites, and randomized link replacement to explore new configurations.

\begin{algorithm}[t]
\caption{Iterative Inter-Plane Link Optimization}
\label{alg:isl_optimization}
\begin{algorithmic}[1]
\State \textbf{Input:} Satellite set $S$, link distance limit $d_{\max}$, link cap $L_{\text{inter}}$, total iterations $n$, degree reinforcement interval $K$, number of satellites modified per iteration $m$
\State \textbf{Output:} Optimized inter-plane link graph $G^*$
\vspace{0.5em}
\State $G^* \gets$ \Call{CreateInitialGraph}{$S$, $d_{\max}$}
\State Evaluate $G^*$ to obtain: worst shortest-path cost $H^*$, avg shortest-path cost $\bar{H}^*$, stability ratio $\sigma^*$
\For{$i = 1$ to $n$}
    \State $G' \gets$ deepcopy($G^*$)
    \If{$i \bmod K = 0$} \Comment{Degree Reinforcement Phase}
        \State Identify underlinked satellites $U$ where inter-degree $< L_{\text{inter}}$
        \For{each satellite $u \in U$}
            \State Add feasible inter-plane links from $u$ to viable candidates
            \State Ensure each endpoint remains within link budget
        \EndFor
    \Else \Comment{Random Replacement Phase}
        \State Select a random subset $M$ of $m$ satellites.
        \For{each $u \in M$}
            \State Randomly remove one existing inter-plane link $(u,v)$
            \State Add a new feasible link $(u,w)$ to a valid candidate $w$
        \EndFor
    \EndIf
    \State Evaluate $G'$ to obtain: worst shortest-path cost $H'$, avg shortest-path cost $\bar{H}'$, stability ratio $\sigma'$
    \If{$H' < H^*$ \textbf{or} ($H' = H^*$ \textbf{and} $\bar{H}' < \bar{H}^*$) \textbf{or} ($H' = H^*$ \textbf{and} $\bar{H}' = \bar{H}^*$ \textbf{and} $\sigma' > \sigma^*$)}
        \State $G^* \gets G'$, $H^* \gets H'$, $\bar{H}^* \gets \bar{H}'$, $\sigma^* \gets \sigma'$
    \EndIf
\EndFor
\State \Return $G^*$
\end{algorithmic}
\end{algorithm}

The procedure is executed over a fixed number of iterations, and at each step, a modified candidate graph is evaluated and conditionally accepted if it improves performance.

\begin{enumerate}
\item \textbf{Repair Phase (Every K Iterations):}\label{Repair Phase}
 Every $K$-th iteration, known as the ``degree reinforcement phase'', a repair phase is triggered to address connectivity gaps.
The algorithm identifies all ``deficient'' satellites with fewer than $L_{\text{inter}}$ inter-plane links. For each deficient satellite, it attempts to add new links by connecting to valid candidates that have available capacity. To ensure fairness, the list of candidates is shuffled before attempting to add new links. A safeguard mechanism ensures that no satellite exceeds its degree constraint as a result of this process. This phase corrects structural weaknesses and prevents sparsely connected nodes from becoming communication bottlenecks.

\item \textbf{Random Placement Phase (Default Iteration):}\label{Random Phase}
In all other iterations, the algorithm performs a random replacement step to explore the solution space. A random subset of satellites is selected, and for each, one of its existing inter-plane links, if any, is randomly removed. The algorithm then attempts to form a new inter-plane link with a different, randomly selected valid candidate, respecting all constraints. This phase introduces topological perturbations, allowing the search to escape local minima and discover more efficient configurations.

\item \textbf{Evaluation and Acceptance: }\label{Eval Phase}
After each iteration, the newly modified graph, $G'$, is evaluated and compared against the best-found graph so far, $G^*$. A candidate graph is accepted based on an ordered comparison of three performance metrics. The new graph $G'$ replaces $G^*$ if:
\begin{enumerate}
    \item It has a smaller 
    diameter;
    or
    \item It has the same diameter but a smaller average 
    shortest-path cost;
    or
    \item Both the diameter and average 
    shortest-path cost
    are the same, but it has a higher fraction of links that remain stable over an orbital period.
\end{enumerate}
This multi-objective acceptance criterion ensures that the optimization greedily improves upon the primary objective of reducing diameter, while using secondary metrics for tie-breaking to enhance overall network performance and stability.

\end{enumerate}

As described earlier, the optimization algorithm is guided by two variants with distinct link feasibility rules. In the snapshot-only variant, the set of valid inter-plane candidates for each satellite is larger, as links need only be feasible at a single instant. This greater topological flexibility allows link stability to be used as a secondary metric to guide the optimization. In contrast, the viability-constrained variant only permits links that remain geometrically valid over a given time (e.g., an orbital period). 
In this stricter model, link stability is guaranteed by construction and therefore does not serve as a separate optimization criterion.

This adaptive local search method, detailed in Algorithm~\ref{alg:isl_optimization}, efficiently navigates the solution space to find topologies that balance a low network diameter with long-term link stability.

\begin{figure*}
    \centering
    \begin{minipage}{\textwidth}
        \centering
        \includegraphics[width=0.85\textwidth]{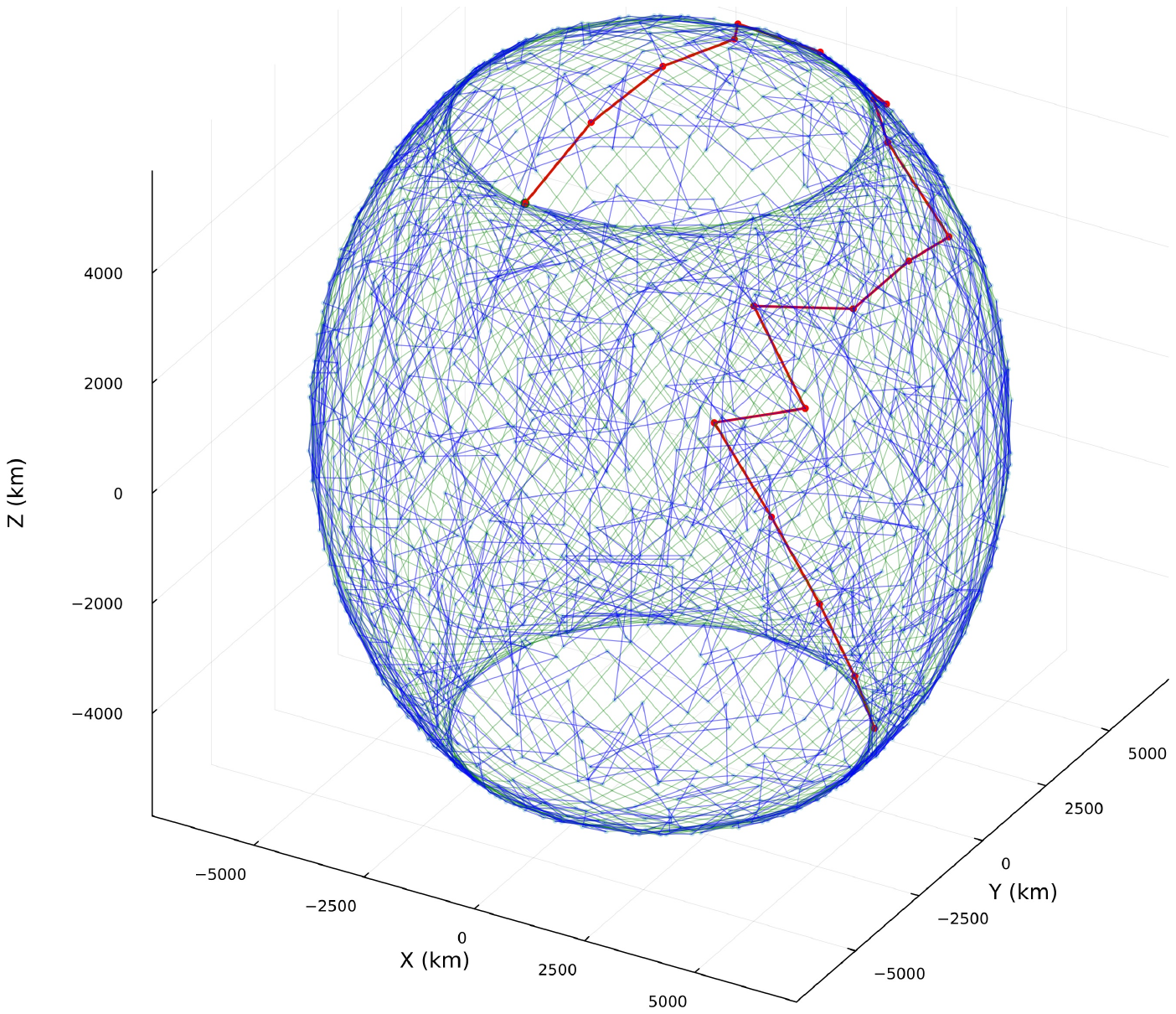}
        \caption{A 3D visualization of the optimized LEO constellation in the viability-constrained model. Intra-plane links are shown in green, while inter-plane links are shown in blue. One of the longest shortest paths is highlighted in red, illustrating the network diameter.}
        \label{fig:constellation}
    \end{minipage}

    \vspace{1em}

    \begin{minipage}{\textwidth}
        \centering
        \includegraphics[width=0.83\textwidth]{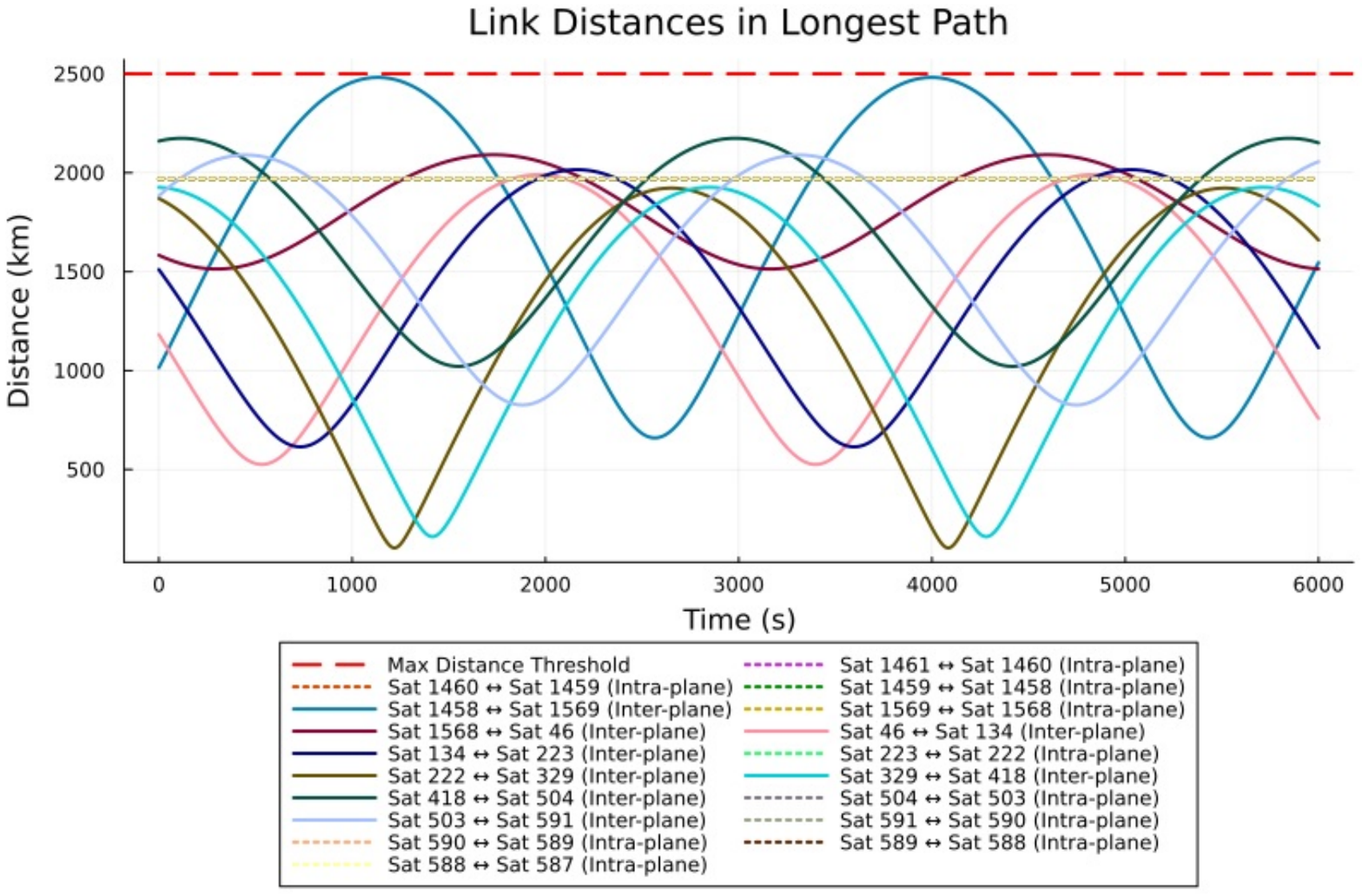}
        \caption{Link distances over time for the highlighted longest shortest path. Intra-plane links (dashed lines) maintain a constant distance, while inter-plane links (solid lines) vary. All links remain below the $d_{\max}$ threshold.}
        \label{fig:link_distances}
    \end{minipage}
\end{figure*}

\section{Simulation \& Results}\label{Simulation & Results}

We evaluate the proposed ISL optimization algorithm on a simulated Walker-Delta constellation. To capture the network asymmetry inherent in practical deployments, where satellites in different orbital planes exhibit slight positional variations, we introduce randomized angular phase offsets $\phi_i \in [0, \phi_{\max}]$ for each orbital plane.

This work focuses on an unweighted graph model, where the objective is to minimize the network diameter, defined here as the shortest-path hop count. Reducing the diameter in terms of hops directly lowers the worst-case time required to broadcast a message across the entire constellation.

The simulation is conducted using the parameters specified in 
Table~\ref{tab:constellation_params}. These parameters are inspired by Starlink Shell 1 configurations~\cite{fcc21}.
The optimization runs for $n=300$ iterations per trial. A degree reinforcement phase is triggered every $K=15$ iterations, while in all other steps, the random replacement phase modifies a subset of $m=20$ satellites.
This entire process is repeated across 50 randomized initializations to ensure robustness.

\begin{table}[h]
\centering
\caption{Constellation and System Parameters}
\label{tab:constellation_params}
\renewcommand{\arraystretch}{1.2}
\begin{tabular}{|c|c|}
\hline
\textbf{Parameter} & \textbf{Value} \\
\hline
Orbital planes ($N_p$) & 72 \\
Satellites per plane ($N_s$) & 22 \\
Total satellites ($N$) & 1584 \\
Orbital altitude ($h$) & 550 km \\
Earth radius ($R_E$) & 6371 km \\
Orbital inclination ($\theta$) & $53^\circ$ \\
Intra-plane links per sat. & 2 \\
Max inter-plane links per sat. ($L_{\text{inter}}$) & 2 \\
Max ISL distance ($d_{\max}$) & 2500 km \\
Phase offset range per plane ($\phi_{\text{max}}$) & $[0, \pi/4]$ (uniform) \\
\hline
\end{tabular}
\end{table}

\begin{table*}[t]
\centering
\caption{Performance Comparison Across Topologies. A viability constraint requires that all inter-plane links remain within $d_{\max}$ over a full orbital cycle.}
\label{tab:results_summary}
\renewcommand{\arraystretch}{1.2}
\begin{tabular}{|l|c|c|c|c|}
\hline
\textbf{Configuration} & \textbf{Max Hop Count} & \textbf{Avg Max Hop} & \textbf{Stable ISL \%} & \textbf{Worst-Case Delay (ms)} \\
\hline
\multicolumn{5}{|c|}{\textbf{With Viability Constraint}} \\
\hline
Optimized (Sparse) & 18 & 17.01 & 100\% & $\approx 125$ \\
Lower Bound (Dense) & 13 & 12.05  & 100\% & $\approx 90$ \\
\hline
\multicolumn{5}{|c|}{\textbf{Without Viability Constraint (Snapshot-Only)}} \\
\hline
Optimized (Sparse) & 13 & 12.02 & 63.84\% & $\approx 90$ \\
Lower Bound (Dense) & 10 & 9.91  & 34.57\% & $\approx 75$ \\
\hline
\end{tabular}
\end{table*}

The performance of each graph is evaluated through an all-source shortest-path analysis to determine its key latency metrics. A breadth-first search (BFS) is executed from each satellite as a source node to find its longest shortest-path distance in hops~\cite{kozen1992depth}. The network diameter and average maximum hop count are then derived from this complete set of measurements.

To measure a topology's long-term stability, we evaluate each inter-plane ISL over a simulation period of $T=6000$~seconds---approximately one full orbital period for satellites at a 550~km altitude. The stability is quantified as the fraction of links that continuously satisfy both the maximum distance ($d_{\max}$) and line-of-sight constraints for this entire duration.

We evaluate each topology using three metrics:

\begin{itemize}
    \item \textbf{Maximum hop count:} The longest shortest-path (in hops) between any pair of satellites. This reflects the worst-case end-to-end latency
    \item \textbf{Average maximum hop count:} The average, over all satellites, of the maximum hop distance from each node to any other. This reflects the typical worst-case broadcast efficiency.
    \item \textbf{Stable ISL fraction:} The fraction of inter-plane links that remain viable (i.e., within range and unobstructed) 
    over an orbital period, capturing topological stability.
\end{itemize}

\subsection{Performance Benchmarks}
To contextualize the performance of our optimized topologies, we establish both theoretical and empirical lower bounds.

The \textbf{theoretical lower bound} is derived from pure geometry. The longest path in the constellation is the geodesic arc between two antipodal satellites ($\approx 21,743$~km). Assuming each hop could span the maximum distance of $d_{\max}=2500$~km, the minimum possible hop count is $D_{\min} = \left\lceil \frac{21{,}743}{2{,}500} \right\rceil = 9$. Complementing this, the hard physical limit on one-way propagation delay over this distance is approximately $\frac{21{,}743 \,\text{km}}{3 \times 10^{5} \,\text{km/s}} \;\approx\; 72.5 \,\text{ms}$. This ideal is unattainable due to orbital mechanics but serves as an absolute structural limit.

To create a more practical benchmark, we derive an \textbf{empirical lower bound} by simulating a ``dense'' topology. For this, we substantially relax the hardware constraint by significantly increasing the maximum number of inter-plane links ($L_{\text{inter}}$) per satellite, setting it to a value greater than the average number of candidate neighbors per satellite. This effectively removes the link budget limitation, making the topology constrained only by geometry. While this dense configuration provides a useful performance floor, its resulting diameter may not be achievable in a sparse network where $L_{\text{inter}}=2$.

\subsection{Snapshot-Only Variant Results}
In the snapshot-only variant, where links need only be feasible at a single instant, our algorithm produces a sparse topology with a maximum hop count of 13 and an average maximum hop count of 12.02 (Table~\ref{tab:results_summary}). While this configuration offers the lowest diameter among the sparse topologies, its drawback is stability: only 63.84\% of its inter-plane links remain viable over a full orbital period.

For comparison, the dense lower-bound topology under this same variant achieves a diameter of 10, closely approaching the theoretical limit of 9. This indicates that our sparse optimization achieves a diameter within 30\% of the empirical best-case scenario, though at the cost of significant link instability.

\subsection{Viability-Constrained Variant Results}
In the viability-constrained variant, where all links must remain stable for an entire orbital period, the trade-off shifts from performance to robustness. The optimized sparse topology achieves a maximum hop count of 18 and an average of 17.01, with 100\% link stability as required by construction (Table~\ref{tab:results_summary}). 

The corresponding dense lower-bound topology in this variant achieves a diameter of 13. This demonstrates that even with unlimited laser terminals, geometric constraints and the requirement for long-term stability impose a hard limit on network diameter that is considerably higher than the theoretical ideal. Our sparse configuration's diameter of 18 is reasonably close to this limit.

\subsection{Topology Visualization and Link Stability}
To visualize the structure and stability of the optimized network, we generate two key plots. Figure~\ref{fig:constellation} shows a 3D rendering of the satellite constellation around the Earth. The links forming one of the longest shortest paths are highlighted, providing a clear visual representation of the network's diameter in the context of the global topology.

Figure~\ref{fig:link_distances} analyzes the stability of this specific path over the full simulation period of $T=6,000$~seconds. The distance of each link in the path is plotted against time. The intra-plane links appear as flat, horizontal lines, confirming their constant separation distance. In contrast, the inter-plane links appear as curves, showing how their distances vary as the satellites move along their orbits. A horizontal line indicates the maximum allowable distance ($d_{\max}$), visually confirming that all links in this path remain viable throughout the simulation.

\subsection{Discussion}
The results, summarized in Table~\ref{tab:results_summary} and visualized in Figure~\ref{fig:constellation}, clearly illustrate the fundamental trade-off between latency performance and operational stability. The snapshot-only approach yields a lower-diameter network (13 hops) by exploiting transient, short-term links. However, this comes at the cost of high link churn (only 63.84\% stability), which would necessitate frequent topology updates and complex dynamic routing protocols, potentially negating the latency benefits.

In contrast, the viability-constrained approach produces a completely stable topology with a higher diameter (18 hops). While its worst-case latency is greater, this configuration offers significant operational simplicity and predictability, as confirmed by the link stability plot in Figure~\ref{fig:link_distances}. Such a static or semi-static topology is well-suited for centrally managed mega-constellations where routing tables can be pre-computed and link reconfiguration is minimized. The choice between these two design philosophies ultimately depends on the specific requirements of the LEO system, balancing the need for low latency against the tolerance for network dynamism and complexity.

\section{Conclusions}\label{Conclusion}
This paper presents a topology optimization framework for minimizing end-to-end communication latency in large-scale LEO satellite constellations. We formulate the design task as a minimax problem that seeks to reduce the network diameter---the maximum shortest-path distance between any pair of satellites---under realistic geometric and hardware constraints.

Our approach operates on static snapshots of satellite positions and supports both snapshot-only and viability-constrained variants. Simulation results demonstrate that enforcing long-term link stability yields robust, deployable topologies with consistent performance across orbital cycles. In contrast, the snapshot-only variant enables lower latency through more aggressive link selection, though at the cost of link churn and reduced stability. This trade-off highlights the flexibility of our framework to adapt to different system priorities, whether emphasizing performance or operational robustness.

The method is particularly applicable to centralized architectures such as a mega-constellation, 
where static or semi-static ISL configurations, cut-through switching, and global route precomputation are feasible. By explicitly modeling link constraints and propagation delay, our framework captures critical dynamics in LEO networks and provides a practical path toward latency-optimized deployments.

Future work may extend this framework to incorporate distance-weighted path metrics, time-aware routing strategies, or hierarchical multi-shell architectures. Additional gains may be realized through the integration of higher-altitude relay layers or dynamic link adaptation mechanisms, enabling scalable, low-latency connectivity across increasingly complex satellite networks.

\acknowledgements
This work was supported in part by NSF grants Nos.~2132700 (SpectrumX) and 2434044.

\bibliographystyle{IEEEtran}
\bibliography{refs}

\end{document}